\documentclass[sigconf]{acmart}
\renewcommand\footnotetextcopyrightpermission[1]{} % removes footnote with conference information in first column
\def\BibTeX{{\rm B\kern-.05em{\sc i\kern-.025em b}\kern-.08emT\kern-.1667em\lower.7ex\hbox{E}\kern-.125emX}}

\setcopyright{none}
\begin{document}

%
% The "title" command has an optional parameter, allowing the author to define a "short title" to be used in page headers.
\title{Sequential modeling of Sessions using Recurrent Neural Networks for Skip Prediction}

%
% The "author" command and its associated commands are used to define the authors and their affiliations.
% Of note is the shared affiliation of the first two authors, and the "authornote" and "authornotemark" commands
% used to denote shared contribution to the research.
\author{Sainath Adapa}
\affiliation{\institution{FindHotel}}

%
% By default, the full list of authors will be used in the page headers. Often, this list is too long, and will overlap
% other information printed in the page headers. This command allows the author to define a more concise list
% of authors' names for this purpose.

%
% The abstract is a short summary of the work to be presented in the article.
\begin{abstract}
Recommender systems play an essential role in music streaming services, prominently in the form of personalized playlists. Exploring the user interactions within these listening sessions can be beneficial to understanding the user preferences in the context of a single session. In the Spotify Sequential Skip Prediction Challenge\footnote{https://www.crowdai.org/challenges/spotify-sequential-skip-prediction-challenge}, WSDM, and Spotify are challenging people to understand the way users sequentially interact with music. We describe our solution approach in this paper and also state proposals for further improvements to the model. The proposed model initially generates a fixed vector representation of the session, and this additional information is incorporated into an Encoder-Decoder style architecture. This method achieved the seventh position in the competition\footnote{Team name: Sainath A}, with a mean average accuracy of 0.604 on the test set. The solution code is available on GitHub\footnote{https://github.com/sainathadapa/spotify-sequential-skip-prediction}.
\end{abstract}



%
% Keywords. The author(s) should pick words that accurately describe the work being
% presented. Separate the keywords with commas.

%
% This command processes the author and affiliation and title information and builds
% the first part of the formatted document.
\maketitle

\section{Introduction}
In many of the music streaming services such as Spotify\footnote{https://www.spotify.com}, personalized music recommendation systems play a prominent role. These recommendation systems allow the user to listen to suggested music based on a particular song, the user's mood, time or location. The vast amount of available music and diverse interests exhibited by various users, as well as by the same user during different situations pose considerable challenges to such systems. As part of WSDM Cup 2019\footnote{http://www.wsdm-conference.org/2019/wsdm-cup-2019.php}, the Spotify Sequential Skip Prediction Challenge mainly explores the sequential nature of user interactions during music listening sessions.

Spotify has provided 130 million listening sessions for training for this challenge. Another 30 million sessions are provided as the test set\cite{brost2019music}. Each session is divided into two nearly equal halves, with the information about tracks available for both halves of a session. However, the user interaction features are available only for the first half\footnote{Also referred to as \textit{session log features} in this document}. The task is to predict if the user skipped any of the tracks in the second half.

The length of each session varies from 10 to 20 tracks. This means the model has to predict skipping behavior for five tracks for the shortest sessions, and ten tracks for the longest. Metadata such as duration, release year, and US popularity estimate is provided for every track. Also, audio features such as acousticness, tempo, loudness are provided. For each track that the user was presented within the session, interactions such as seek forward/backward, short/long pause before play are available. Finally, session information such as the time of the day, date, premium user or not, context type of playlist is present.

In the dataset, skipping behavior is classified into four types:
\begin{enumerate}
    \item \textit{skip\_1}: Boolean indicating if the track was only played very briefly
    \item \textit{skip\_2}: Boolean indicating if the track was only played briefly
    \item \textit{skip\_3}: Boolean indicating if most of the track was played
    \item \textit{not\_skipped}: Boolean indicating that the track was played in its entirety
\end{enumerate}
The objective of the challenge is limited to predicting just the \textit{skip\_2} behavior.

Tables 1 to 3 present distributions for some of the features in the dataset\footnote{Owing to the size of the dataset, a random sample of data was used to calculate the distributions for Table 2 and 3}. From Table 2, it can be inferred that the \textit{skip\_2} variable is fairly balanced between the true and false classes. We can also observe, from Table 3, that the skipping behavior remained consistent from track 4.

The test set distributions of features were found to be very similar to that of the training set. Hence, neither sampling nor other modifications were made to the training set before training the model. 

\begin{table*}
\caption{Distribution of number of tracks in sessions}
\begin{tabular}{l|l|l|l|l|l|l|l|l|l|l|l|}
\cline{2-12}
Number of tracks & 10    & 11    & 12    & 13    & 14    & 15    & 16    & 17    & 18    & 19    & 20     \\ \cline{2-12} 
Percentage       & 8.8\% & 7.7\% & 6.7\% & 5.9\% & 5.2\% & 4.5\% & 4.0\% & 3.5\% & 3.1\% & 2.8\% & 47.7\% \\ \cline{2-12} 
\end{tabular}
\end{table*}

\begin{table}
\caption{Overall skipping behavior}
\begin{tabular}{|l|l|}
\hline
Type         & True percentage \\ \hline
\textit{skip\_1}      & 41.52\%           \\ \hline
\textit{skip\_2}      & 50.89\%           \\ \hline
\textit{skip\_3}      & 63.86\%           \\ \hline
\textit{not\_skipped} & 34.41\%          \\ \hline
\end{tabular}
\end{table}

\begin{table*}
\caption{"skip\_2 = true" percentage by session position}
\begin{tabular}{|l|l|l|l|l|l|l|l|l|l|l|l|l|l|l|l|l|l|l|l|}
\hline
1    & 2    & 3    & 4    & 5    & 6    & 7    & 8    & 9    & 10   & 11   & 12   & 13   & 14   & 15   & 16   & 17   & 18   & 19   & 20   \\ \hline
37.4 & 46.7 & 49.4 & 51.6 & 52.4 & 53.2 & 53.1 & 53.0 & 52.1 & 50.3 & 50.7 & 51.3 & 51.7 & 52.2 & 52.5 & 53.0 & 53.3 & 53.5 & 53.6 & 53.6 \\ \hline
\end{tabular}
\end{table*}

\section{Related Work}
Deep learning based recommendation methods have been extensively studied during recent years \cite{zhang2017deep}. Session-based music recommender systems specifically try to infer user's preferences and context using information from a single session. This differs from \textit{session-aware} systems which use previous interactions of the same user in the recommendation process \cite{quadrana2018sequence}. The current task shares many aspects of session-based music recommender systems. Hence, understanding the approaches that are employed to such systems is useful for the current task. A survey and evaluation of various approaches to Session-based recommendation systems was presented in \cite{ludewig2018evaluation}. Recurrent Neural Networks (RNNs) have been shown to work exceedingly well with sequential modeling tasks \cite{chung2014empirical}. As such, in \cite{hidasi2015session}, a new architecture named GRU4REC that employs Gated Recurrent Units (GRUs) was proposed to predict the probability of subsequent events given a session beginning. A data augmentation technique that improves upon \cite{hidasi2015session} via sequence pre-processing was proposed in \cite{tan2016improved}. To model the changes in user behavior based on context, a new recurrent architecture was proposed in \cite{smirnova2017contextual}.

\section{Approach}
\subsection{Modified Encoder-Decoder Architecture}
In 2014, \citet{cho2014learning} proposed the Encoder-Decoder architecture, that consisted of two recurrent neural networks (RNN). The Encoder RNN strives to encode the input sequence into a fixed length representation, and from this representation, the Decoder RNN generates a correct, variable length target sequence. The architecture was proposed for the statistical machine translation, for which it was shown to be a significant improvement over previous methods.

The current task of sequential skip prediction shares the variable length and the sequential dependency aspects of the statistical machine translation. However, differing significantly from the statistical machine translation task, the current data set contains information about the tracks in the second half. There is a direct one-to-one correspondence between the track and the output skip prediction. In the following sections, We describe a modified Encoder-Decoder architecture that takes advantage of the unique characteristics of the present dataset.

\subsection{Base transformation of input data}
The raw input feature vector space might not be an ideal representation for processing by the Long Short-Term Memory (LSTM) cells in the model. Hence, a single Fully-Connected (FC) layer was used to transform the user interaction features and other metadata about the session into a higher dimensional representation. Similarly, a separate FC layer was used to process the acoustic and other metadata associated with tracks. Note that the same FC layer mutates tracks from both the first and the second half of the session. Both the FC layers were equipped with the rectification (ReLU) non-linearity. Only the transformed features were used subsequently in the model. 

\subsection{Compact representation of the session}
As shown by \citet{cho2014properties}, Encoder-Decoder architectures exhibit weakness in handling long sentences. The fixed-length vector representation that is transferred from encoder to decoder may not have enough capacity to encode the complicated relationships within the data. Many solutions have been proposed to mitigate this issue \cite{luong2015effective} \cite{vaswani2017attention}. In the current approach, a compact representation of the session is generated to preserve information for use by Decoder. This fixed length vector is then concatenated with each track's features, before being fed into the Encoder or Decoder.

The track features corresponding to the first half of the session were concatenated with the first half's user interaction features. This was then fed into a Bi-directional Long Short-Term Memory (BiLSTM) network. The final output from this Bi-directional LSTM can be considered to contain aspects such as overall user behavior in the session and user behavior with respect to specific track features within the session.

During the previous computation, we have only considered the first half of the session. However, we also have the track information for the second half of the session. Usage of track information from both the first and second halves can lead to a better understanding of the nature of the playlist. Hence a second BiLSTM layer was used to transform the long sequence of all the tracks within the session into a fixed length vector.

The combined output from both the Bi-directional LSTMs can now be considered as a compact representation of all the information that is available to the model as input.

\begin{figure}[h]
\caption{Computing a fixed vector representation of the session}
\centering
\includegraphics[scale=0.5]{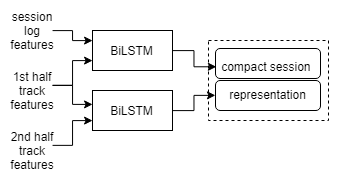}
\end{figure}

\subsection{Encoder}
At every track's prediction, the Decoder now has easy access to a representation of the session from the input. Hence, the only goal of the Encoder currently is to create the initial state for the decoder. The Encoder consists of an FC layer and a subsequent BiLSTM layer.

\begin{figure}[h]
\caption{Encoder architecture}
\centering
\includegraphics[scale=0.5]{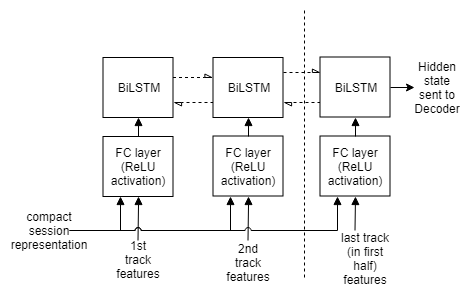}
\end{figure}

\subsection{Decoder}
The track features of the second half along with fixed vector representation of the session is first sent to an FC layer with ReLU non-linearity. The purpose of this layer can be seen as a context setting mechanism - the track features are being transformed using the current session as the context. Note that the weights of this FC layer are shared with that of the FC layer from the Encoder. 

The output from the previous layer is now sent to a BiLSTM layer. For the very first track in the second half, the final state from the BiLSTM layer of the encoder is used as the hidden state for this layer.

Skipping behavior exhibited by the user during the previous track is a significant predictor for the current track. Hence, the output from the Decoder for the previous track is combined with the output from the previous layer. In case of the first track, the actual \textit{skip\_2} value of the last track in the first half is used. This combined vector is sent to an LSTM layer. Finally, the output from the LSTM layer is fed into a Fully Connected layer with Sigmoid non-linearity to generate the prediction.

\begin{figure}[h]
\caption{Decoder architecture}
\centering
\includegraphics[scale=0.5]{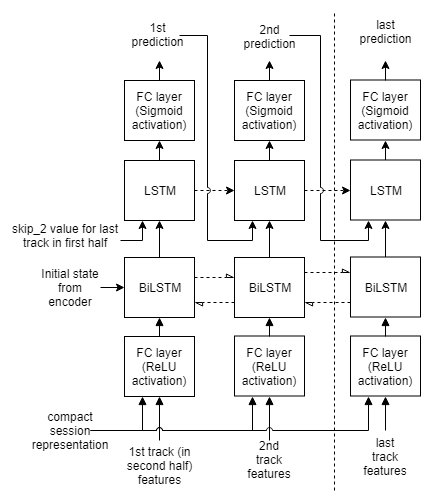}
\end{figure}

\section{Model training}
\subsection{Evaluation metric}
For evaluation, Mean Average Accuracy (MAA) was selected as the primary metric for the challenge. Here, the average accuracy is defined as
\[ AA = \sum_{i=1}^T \frac{A(i)L(i)}{T} \]
where:
\begin{itemize}
    \item \textit{T} is the number of tracks to be predicted for the given session
    \item \textit{A(i)} is the accuracy at position \textit{i} of the sequence
    \item \textit{L(i)} is the Boolean indicator for if the \textit{i}'th prediction was correct
\end{itemize}

The motivation for the metric is the notion that the immediate track's prediction is most important. As a tie-breaking secondary metric, the average accuracy of the first prediction was used.

\subsection{Loss function and weights}

\begin{table}
\caption{Average Accuracy values for one simulated sample of length 5}
\begin{tabular}{|l|l|l|}
\cline{1-1} \cline{2-2} \cline{3-3}
Ground truth sequence & Predicted sequence & AA   \\ \hline
1, 1, 1, 1, 1         & 0, 1, 1, 1, 1      & 0.543 \\ \hline
1, 1, 1, 1, 1         & 1, 0, 1, 1, 1      & 0.643 \\ \hline
1, 1, 1, 1, 1         & 1, 1, 0, 1, 1      & 0.710  \\ \hline
1, 1, 1, 1, 1         & 1, 1, 1, 0, 1      & 0.760  \\ \hline
1, 1, 1, 1, 1         & 1, 1, 1, 1, 0      & 0.800   \\ \hline
\end{tabular}
\end{table}

Table 4 shows Average Accuracy values for various predictions on a sequence of length 5. As illustrated by the table, a wrong prediction of the first track decreases the AA value by 0.457 whereas an incorrect prediction of the last track would reduce the AA by just 0.2. One way to interpret this is that the first prediction is (0.457/0.2 =) 2.285 times as important as the last prediction in this example. Incorporating this information into the loss function is useful. Hence, session positions were allocated weights in proportion to the decrease in Average Accuracy value when the prediction for that session position alone is incorrect. Log Loss with weights assigned for each session position was used for optimizing the parameters of the model.

\subsection{Training}
Due to time and resource constraints, we were not able to train the model on the whole data set. Three mutually exclusive sets of sessions were sampled from the data - Training, Validation, and Test. The Model was trained on the training set, using the validation set to determine the stopping point. The parameters were tuned until the loss on validation set plateaued. The third sample - Test set was used to perform a local evaluation of the model.

\subsection{Results}
Before the results of the proposed method are presented, we introduce the result of a baseline model. The baseline model uses the skipping behavior of the last track in the first half as the prediction for all the tracks in the second half. This model scored 0.537 on MAA and 0.742 on First prediction accuracy. Predictions from the proposed model in this paper resulted in an MAA score of 0.604 and First Prediction accuracy of 0.792 on the hold-out set, thus achieving the seventh position in the competition.

\section{Future work}
Only around 20\% of the data was used for training the model because of resource constraints. Instead, training the model on 80\% of the data (leaving 20\% of the sessions for validation and test sets) might improve accuracy. 

During the exploration phase, a Random Forest (RF) model was built to predict the first track of the second half. Another RF model was trained to predict the last track in the second half. This model was built using the session log features until the last track, which is unlike the setup of this challenge where session log features are only available for the first half. Both the models achieved similar levels of accuracy. This reveals that fundamentally, user behavior during the end of the second half is not much more variable (and thus not harder to predict) than during the beginning of the second half. If complete data is available until the previous track, then any session position can be predicted with reasonable accuracy.

In the absence of complete data, we can try to generate predictions for missing features by building a model that predicts those features. Hence if a model was trained to predict all types of user interactions (\textit{skip\_1}, \textit{2}, and \textit{3}, \textit{hist\_user\_behavior\_reason\_start}, and other remaining features), it is possible that the resulting model might be better at predicting \textit{skip\_2}. Similar to the proposed model in this paper, this model can use the previous time step's predictions while predicting the current time step.

Another option is to employ transfer learning. Transfer learning has been useful in many areas, with one prominent example being the use of ImageNet-trained models\cite{huh2016makes}. As described in the previous paragraph, we can build and train a model to predict all the user interaction features available in the data. Such a model theoretically would have inferred more aspects of the user behavior than a model that is solely trained for \textit{skip\_2} prediction. We can then fine-tune the top layers of this model specifically for \textit{skip\_2} resulting in better \textit{skip\_2} prediction accuracy.

%
% The acknowledgments section is defined using the "acks" environment (and NOT an unnumbered section). This ensures
% the proper identification of the section in the article metadata, and the consistent spelling of the heading.
\begin{acks}
We want to thank WSDM, Spotify, and CrowdAI for organizing the challenge. Special thanks to Google for providing the coupons for using Google Cloud Compute resources. Considering the large size of the dataset, the coupons were especially helpful.
\end{acks}

%
% The next two lines define the bibliography style to be used, and the bibliography file.
\bibliographystyle{ACM-Reference-Format}
%\bibliography{sample-base}

\begin{thebibliography}{13}

%%% the Web address.

\ifx \showCODEN    \undefined \def \showCODEN     #1{\unskip}     \fi
\ifx \showDOI      \undefined \def \showDOI       #1{#1}\fi
\ifx \showISBNx    \undefined \def \showISBNx     #1{\unskip}     \fi
\ifx \showISBNxiii \undefined \def \showISBNxiii  #1{\unskip}     \fi
\ifx \showISSN     \undefined \def \showISSN      #1{\unskip}     \fi
\ifx \showLCCN     \undefined \def \showLCCN      #1{\unskip}     \fi
\ifx \shownote     \undefined \def \shownote      #1{#1}          \fi
\ifx \showarticletitle \undefined \def \showarticletitle #1{#1}   \fi
\ifx \showURL      \undefined \def \showURL       {\relax}        \fi
% invisible to TeX
\providecommand\bibfield[2]{#2}
\providecommand\bibinfo[2]{#2}
\providecommand\natexlab[1]{#1}
\providecommand\showeprint[2][]{arXiv:#2}

\bibitem[\protect\citeauthoryear{Brost, Mehrotra, and Jehan}{Brost
  et~al\mbox{.}}{2019}]%
        {brost2019music}
\bibfield{author}{\bibinfo{person}{Brian Brost}, \bibinfo{person}{Rishabh
  Mehrotra}, {and} \bibinfo{person}{Tristan Jehan}.}
  \bibinfo{year}{2019}\natexlab{}.
\newblock \showarticletitle{The Music Streaming Sessions Dataset}. In
  \bibinfo{booktitle}{\emph{Proceedings of the 2019 Web Conference}}. ACM.
\newblock


\bibitem[\protect\citeauthoryear{Cho, Van~Merri{\"e}nboer, Bahdanau, and
  Bengio}{Cho et~al\mbox{.}}{2014a}]%
        {cho2014properties}
\bibfield{author}{\bibinfo{person}{Kyunghyun Cho}, \bibinfo{person}{Bart
  Van~Merri{\"e}nboer}, \bibinfo{person}{Dzmitry Bahdanau}, {and}
  \bibinfo{person}{Yoshua Bengio}.} \bibinfo{year}{2014}\natexlab{a}.
\newblock \showarticletitle{On the properties of neural machine translation:
  Encoder-decoder approaches}.
\newblock \bibinfo{journal}{\emph{arXiv preprint arXiv:1409.1259}}
  (\bibinfo{year}{2014}).
\newblock


\bibitem[\protect\citeauthoryear{Cho, Van~Merri{\"e}nboer, Gulcehre, Bahdanau,
  Bougares, Schwenk, and Bengio}{Cho et~al\mbox{.}}{2014b}]%
        {cho2014learning}
\bibfield{author}{\bibinfo{person}{Kyunghyun Cho}, \bibinfo{person}{Bart
  Van~Merri{\"e}nboer}, \bibinfo{person}{Caglar Gulcehre},
  \bibinfo{person}{Dzmitry Bahdanau}, \bibinfo{person}{Fethi Bougares},
  \bibinfo{person}{Holger Schwenk}, {and} \bibinfo{person}{Yoshua Bengio}.}
  \bibinfo{year}{2014}\natexlab{b}.
\newblock \showarticletitle{Learning phrase representations using RNN
  encoder-decoder for statistical machine translation}.
\newblock \bibinfo{journal}{\emph{arXiv preprint arXiv:1406.1078}}
  (\bibinfo{year}{2014}).
\newblock


\bibitem[\protect\citeauthoryear{Chung, Gulcehre, Cho, and Bengio}{Chung
  et~al\mbox{.}}{2014}]%
        {chung2014empirical}
\bibfield{author}{\bibinfo{person}{Junyoung Chung}, \bibinfo{person}{Caglar
  Gulcehre}, \bibinfo{person}{KyungHyun Cho}, {and} \bibinfo{person}{Yoshua
  Bengio}.} \bibinfo{year}{2014}\natexlab{}.
\newblock \showarticletitle{Empirical evaluation of gated recurrent neural
  networks on sequence modeling}.
\newblock \bibinfo{journal}{\emph{arXiv preprint arXiv:1412.3555}}
  (\bibinfo{year}{2014}).
\newblock


\bibitem[\protect\citeauthoryear{Hidasi, Karatzoglou, Baltrunas, and
  Tikk}{Hidasi et~al\mbox{.}}{2015}]%
        {hidasi2015session}
\bibfield{author}{\bibinfo{person}{Bal{\'a}zs Hidasi},
  \bibinfo{person}{Alexandros Karatzoglou}, \bibinfo{person}{Linas Baltrunas},
  {and} \bibinfo{person}{Domonkos Tikk}.} \bibinfo{year}{2015}\natexlab{}.
\newblock \showarticletitle{Session-based recommendations with recurrent neural
  networks}.
\newblock \bibinfo{journal}{\emph{arXiv preprint arXiv:1511.06939}}
  (\bibinfo{year}{2015}).
\newblock


\bibitem[\protect\citeauthoryear{Huh, Agrawal, and Efros}{Huh
  et~al\mbox{.}}{2016}]%
        {huh2016makes}
\bibfield{author}{\bibinfo{person}{Minyoung Huh}, \bibinfo{person}{Pulkit
  Agrawal}, {and} \bibinfo{person}{Alexei~A Efros}.}
  \bibinfo{year}{2016}\natexlab{}.
\newblock \showarticletitle{What makes ImageNet good for transfer learning?}
\newblock \bibinfo{journal}{\emph{arXiv preprint arXiv:1608.08614}}
  (\bibinfo{year}{2016}).
\newblock


\bibitem[\protect\citeauthoryear{Ludewig and Jannach}{Ludewig and
  Jannach}{2018}]%
        {ludewig2018evaluation}
\bibfield{author}{\bibinfo{person}{Malte Ludewig} {and}
  \bibinfo{person}{Dietmar Jannach}.} \bibinfo{year}{2018}\natexlab{}.
\newblock \showarticletitle{Evaluation of Session-based Recommendation
  Algorithms}.
\newblock \bibinfo{journal}{\emph{arXiv preprint arXiv:1803.09587}}
  (\bibinfo{year}{2018}).
\newblock


\bibitem[\protect\citeauthoryear{Luong, Pham, and Manning}{Luong
  et~al\mbox{.}}{2015}]%
        {luong2015effective}
\bibfield{author}{\bibinfo{person}{Minh-Thang Luong}, \bibinfo{person}{Hieu
  Pham}, {and} \bibinfo{person}{Christopher~D Manning}.}
  \bibinfo{year}{2015}\natexlab{}.
\newblock \showarticletitle{Effective approaches to attention-based neural
  machine translation}.
\newblock \bibinfo{journal}{\emph{arXiv preprint arXiv:1508.04025}}
  (\bibinfo{year}{2015}).
\newblock


\bibitem[\protect\citeauthoryear{Quadrana, Cremonesi, and Jannach}{Quadrana
  et~al\mbox{.}}{2018}]%
        {quadrana2018sequence}
\bibfield{author}{\bibinfo{person}{Massimo Quadrana}, \bibinfo{person}{Paolo
  Cremonesi}, {and} \bibinfo{person}{Dietmar Jannach}.}
  \bibinfo{year}{2018}\natexlab{}.
\newblock \showarticletitle{Sequence-aware recommender systems}.
\newblock \bibinfo{journal}{\emph{arXiv preprint arXiv:1802.08452}}
  (\bibinfo{year}{2018}).
\newblock


\bibitem[\protect\citeauthoryear{Smirnova and Vasile}{Smirnova and
  Vasile}{2017}]%
        {smirnova2017contextual}
\bibfield{author}{\bibinfo{person}{Elena Smirnova} {and}
  \bibinfo{person}{Flavian Vasile}.} \bibinfo{year}{2017}\natexlab{}.
\newblock \showarticletitle{Contextual sequence modeling for recommendation
  with recurrent neural networks}. In \bibinfo{booktitle}{\emph{Proceedings of
  the 2nd Workshop on Deep Learning for Recommender Systems}}. ACM,
  \bibinfo{pages}{2--9}.
\newblock


\bibitem[\protect\citeauthoryear{Tan, Xu, and Liu}{Tan et~al\mbox{.}}{2016}]%
        {tan2016improved}
\bibfield{author}{\bibinfo{person}{Yong~Kiam Tan}, \bibinfo{person}{Xinxing
  Xu}, {and} \bibinfo{person}{Yong Liu}.} \bibinfo{year}{2016}\natexlab{}.
\newblock \showarticletitle{Improved recurrent neural networks for
  session-based recommendations}. In \bibinfo{booktitle}{\emph{Proceedings of
  the 1st Workshop on Deep Learning for Recommender Systems}}. ACM,
  \bibinfo{pages}{17--22}.
\newblock


\bibitem[\protect\citeauthoryear{Vaswani, Shazeer, Parmar, Uszkoreit, Jones,
  Gomez, Kaiser, and Polosukhin}{Vaswani et~al\mbox{.}}{2017}]%
        {vaswani2017attention}
\bibfield{author}{\bibinfo{person}{Ashish Vaswani}, \bibinfo{person}{Noam
  Shazeer}, \bibinfo{person}{Niki Parmar}, \bibinfo{person}{Jakob Uszkoreit},
  \bibinfo{person}{Llion Jones}, \bibinfo{person}{Aidan~N Gomez},
  \bibinfo{person}{{\L}ukasz Kaiser}, {and} \bibinfo{person}{Illia
  Polosukhin}.} \bibinfo{year}{2017}\natexlab{}.
\newblock \showarticletitle{Attention is all you need}. In
  \bibinfo{booktitle}{\emph{Advances in Neural Information Processing
  Systems}}. \bibinfo{pages}{5998--6008}.
\newblock


\bibitem[\protect\citeauthoryear{Zhang, Yao, and Sun}{Zhang
  et~al\mbox{.}}{2017}]%
        {zhang2017deep}
\bibfield{author}{\bibinfo{person}{Shuai Zhang}, \bibinfo{person}{Lina Yao},
  {and} \bibinfo{person}{Aixin Sun}.} \bibinfo{year}{2017}\natexlab{}.
\newblock \showarticletitle{Deep learning based recommender system: A survey
  and new perspectives}.
\newblock \bibinfo{journal}{\emph{arXiv preprint arXiv:1707.07435}}
  (\bibinfo{year}{2017}).
\newblock


\end{thebibliography}

\end{document}